\documentclass[12pt,preprint]{aastex}
\usepackage{natbib}
\usepackage{epsfig}
\begin{document}

\title{Efficient pseudo-global fitting for helioseismic data}

\author{S. T. Fletcher\altaffilmark{1}}
\affil{Faculty of Arts, Computing, Engineering and Science,
Sheffield Hallam University, Sheffield, UK}
\email{s.fletcher@shu.ac.uk}

\author{W. J. Chaplin and Y. Elsworth}
\affil{School of Physics and Astronomy, University of Birmingham,
Birmingham, UK}

\and

\author{Roger New}
\affil{Faculty of Arts, Computing, Engineering and Science,
Sheffield Hallam University, Sheffield, UK}

\altaffiltext{1}{School of Physics and Astronomy, University of
Birmingham, Birmingham, UK}

\begin{abstract}
Mode fitting or ``peak-bagging" is an important procedure in
helioseismology allowing one to determine the various mode
parameters of solar oscillations. Here we describe a way of reducing
the systematic bias in the fits of certain mode parameters that are
seen when using ``local" fitting techniques to analyse the
sun-as-a-star p-mode power spectrum. To do this we have developed a
new ``pseudo-global" fitting algorithm designed to gain the
advantages of fitting the entire power spectrum, but without the
problems involved in fitting a model incorporating many hundreds of
parameters.

We have performed a comparative analysis between the local and
pseudo-global peak-bagging techniques by fitting the ``limit"
profiles of simulated helioseismic data. Results show that for
asymmetric modes the traditional fitting technique returns
systematically biased estimates of the central frequency parameter.
This bias is significantly reduced when employing the pseudo-global
routine. Similarly, we show that estimates of the background
returned from the pseudo-global routine match the input values much
more closely than the estimates from the local fitting method.

We have also used the two fitting techniques to analyse a set of
real solar data collected by the Global Oscillations at Low
Frequencies (GOLF) instrument on board the ESA/NASA Solar and
Heliospheric Observatory (SOHO) spacecraft. Similar differences
between the estimated frequencies returned by the two techniques are
seen when fitting both the real and simulated data. We show that the
background fits returned by the pseudo-global routine more closely
match the estimate of the background one can infer from
interpolating between fits to the high and low frequency ends of the
p-mode power spectrum.
\end{abstract}


\section{Introduction}

Determining the parameter values of the resonant modes of
oscillation of the Sun is an important process in helioseismology.
Mode frequencies, rotational splittings, lifetimes and amplitudes
can all be used to help identify the conditions of the solar
interior. For example, by employing inversion techniques the mode
frequencies can be used to constrain estimates of the sound speed
and density in the solar interior, while the rotational splitting
parameter can determine the internal rotation rate. The mode
linewidths and amplitudes may be used in order to better understand
possible models of convection within the outer layers of the solar
interior.

Over the years the quality of helioseismic data has improved
significantly due to the length of data sets increasing,
signal-to-noise ratios being improved and more continuous
observations being made, both from ground-based networks and
space-borne missions. This has led to the parameter values being
constrained with increasingly greater precision which in turn has
led to better constraints on our understating of the solar interior.

As the precision of the parameter estimates increases so more subtle
characteristics of the p-mode power spectrum are being uncovered. An
example of this is the discovery that the resonant mode peaks that
make up the p-mode spectrum are actually slightly asymmetric (e.g.,
\citealt{Duvall1993,Chaplin1999}). This supported prior theoretical
predictions (e.g., \citealt{GabrielM1992,GabrielM1995}) and added to
the evidence that acoustic waves are generated within a
well-localized region of the solar interior and are possibly
correlated with the background generated by solar granulation
\citep{Nigam1998,Toutain2006}.

In order to analyse such subtle effects, very accurate and robust
parameter estimates are needed. Any bias in the returned parameters
can lead to confusion about which aspects of the mode
characteristics have true physical significance. Additionally,
quoting very precise estimates of the mode parameters without also
reducing, or at least understanding, possible systematic errors may
lead to significant errors in the solar parameters derived through
inversion techniques.

Methods of determining mode parameters are often referred to as
``peak-bagging" techniques. For low-degree (low-$\ell$)
Sun-as-a-star observations (i.e., observations where the collected
light has been integrated over the entire solar surface), a
traditional method of peak-bagging involves dividing the p-mode
power spectrum into a series of ``fitting windows" centered on the
$\ell$ = 0/2 and 1/3 pairs. The modes are then fitted, pair by pair,
to determine how the mode parameters depend on both frequency
(overtone number) and angular degree without the need to fit the
entire spectrum simultaneously.

The main advantage of a traditional ``pair-by-pair" fitting method
(hereafter abbreviated PPM) is its computational efficiency, since
the number of parameters being varied remains quite small and only a
fraction of the full p-mode spectrum is fitted at any one time.
However, there is a cost for this efficiency since the model used to
fit the data encompasses only those modes whose central frequencies
fall within the fitting window. Therefore, any power from modes
whose central frequencies lie outside this region will not be
accounted for. This imperfect match between the fitting model and
the underlying profile of the data may lead to significant biases in
some of the fitted parameters.

Global fitting approaches that involve fitting the entire spectrum
simultaneously would obviously resolve this issue. However, these
have not been regularly employed due to the large number of
parameters involved in the fitting model. This leads to both long
computing times (especially when fitting long multi-year data sets)
and an increased possibility of premature convergence.

In \cite{Fletcher2008} (hereafter referred to as Paper I) we
described a modified peak-bagging process which retained the
simplicity and efficiency of the PPM but gained the benefits of a
global fitting approach. We refer to this method as a
``pseudo-global" method (hereafter abbreviated PGM). Simulated data
with symmetric mode peaks were used in Paper I and while the PGM
returned less biased estimates for the background, widths and
heights, the fitted frequencies were found to be unbiased whichever
fitting technique was employed. However, recent work has shown that
if the modes are asymmetric then the PPM method may also return
biased estimates of the frequencies (see \citealt{Jimenez2008}).

With this is mind, we have employed a more sophisticated simulator
that creates artificial spectra with asymmetric modes. We have also
modified the PGM in order to correctly account for this added
complexity. Both the PPM and PGM have been tested using simulated
data and then applied to a set of real data collected by the Global
Oscillations at Low Frequencies (GOLF) instrument on board the Solar
and Heliospheric Observatory (SOHO) spacecraft (see
\citealt{GabrielA1995}).

The remainder of this paper is presented in the following order. The
calibration of the GOLF data and the generation of the simulated
data are described in section~2. In section~3 we re-cap the PPM and
explain in detail the algorithm for the new PGM. Finally, in
section~4 we give the results of the fits to both the simulated and
real data. In this paper we concentrate solely on the fits to the
frequency, asymmetry and background parameters.

\section{Data}\label{SecData}

We present the results of testing our new PGM on both simulated and
real GOLF data and compare the results with those returned when
using the PPM. GOLF data were chosen over other real data because
there are relatively long stretches of observations where the
duty-cycle is close to one-hundred percent. This removes any added
complications involved in accounting for an extended window function
(e.g., diurnal gaps in the case of ground based observations) and
hence keeps things simple for this initial test of the PGM.

The GOLF time series used was a 796-day set collected between April
1996 and June 1998. The time series was calibrated according to the
method described in \cite{Garcia2005}, with the data from the two
detectors on board the GOLF instrument being summed in order to
reduce counting noise. This 796-day set was the first extended time
series obtained by the GOLF instrument and was collected by
observing intensity changes in the blue-wing of the chosen
absorption line. The duty cycle for this period was over 99 per-cent
and is thus ideal for testing the modified fitting code in the
presence of very little contamination from window-function effects.

In Paper I Monte Carlo simulations were used to test the reliability
of the fitting procedures. This was done by comparing the average of
the fitted parameters for multiple independent realizations of
simulated data with the known input values used to generate the
data.

Here we take a more efficient approach by creating and fitting
``limit" spectra of the simulated data (i.e., the spectrum one would
obtain in the limit of summing an infinite number of independently
generated spectra). \cite{Toutain2005} showed that comparing the
known input parameters with those returned by fitting the limit
spectrum gives a direct estimate of the bias without the need to fit
many independent realizations of the simulated data. This obviously
leads to a dramatic reduction in the computing time needed to
perform such comparative tests. The one disadvantage of this method
is that it provides only a single value, and its associated
uncertainty, for each fitted parameter. Hence, it is not possible to
analyse the \emph{distribution} of the fitted values as is often
done with a full Monte Carlo treatment. This means, for example,
that checks to make sure that the fitted values are distributed
normally (an assumption made when determining the formal
uncertainties) cannot be performed.

In Paper I the simulated data were created using the
first-generation solarFLAG simulator (see
\citealt{Chaplin1997,Chaplin2006}). The second-generation simulator
is considerably more sophisticated and was developed in order to
incorporate asymmetry in the mode peaks of the power spectrum. This
is achieved by introducing correlations between the excitations of
overtones with the same degree and azimuthal order, $m$, and by
correlating the excitations with the background. A full description
of how this simulator generates a realistic stochastically driven
artificial data set and the method of creating the corresponding
underlying limit profile can both be found in \cite{Chaplin2008}

In order to match the GOLF data the simulated limit spectra had
frequency resolutions and ranges corresponding to a time series of
796 days with a regular 40-s cadence. A full set of low-$\ell$ modes
was included in the simulations with a frequency range 1000 $ \leq
\nu \leq $ 5000 $\mu$Hz and with angular degrees 0 $ \leq \ell \leq
$ 5. The input mode parameters were chosen based on both information
from standard solar models (e.g., see \citealt{Bahcall2005}) and
analysis of GOLF and Birmingham Solar Oscillations Network (BiSON)
data.

Two sources of background need to be considered. The first of these
is a granulation-like component that mimics the solar velocity
continuum. The power spectral density, $n$, as a function of
frequency, $\nu$, may be described by a single-term power-law model
\citep{Harvey1985}, i.e.,
\begin{equation}
n(\nu) = \frac{2\sigma^2\tau}{1+(2\pi\nu\tau)^2} \label{Harvey}
\end{equation}
where $\tau$ and $\sigma$ represent the time constant and standard
deviation respectively of the granulation signal. The value of these
constants were chosen so as to match the background profile of the
real GOLF data as closely as possible. Secondly, white noise with a
Gaussian distribution is added to simulate the uncertainty in the
velocity measurements caused by photon shot noise.

For the purposes of this paper correlations with the background were
handled in two different ways. For one limit spectrum the excitation
profiles were correlated directly with the granulation-like
background, and in the other, the excitation profiles were
correlated with a flat background function and an uncorrelated
granulation-like background was added in later. The reason for this
was to investigate different sources of asymmetry of the mode peaks.
If asymmetry in the real data is caused (at least in part) by
correlation between the background and the excitations then the
first case should provide the best model of this. Whereas, if the
asymmetry is generated in other ways, then the flat background
correlated data may provide a better analogue of the real p-mode
spectrum. Of course it is most likely that the asymmetry is actually
caused by a combination of effects.

\section{Fitting Techniques}\label{SecTechniques}

In this section we describe in detail the PGM routine. The PPM has
already been well documented elsewhere (e.g., see
\citealt{Chaplin1999}), but as it makes up the first step in
employing the PGM we first highlight a number of specific points
where its application may differ from that in the literature.

For the PPM power spectra are split into a series of fitting windows
centered on both $\ell$=0/2 and $\ell$=1/3 pairs. Within each
fitting window mode peaks are modeled using an asymmetric Lorentzian
profile of the form:
\begin{equation}
P(\nu) = A \frac{1+2bx}{1+x^2} \label{AsyLorentz}
\end{equation}
where $x = (\nu-\nu_0)/\Gamma$, and $\nu_0$, $\Gamma$, $A$ and $b$
are the mode central frequency, linewidth, amplitude and asymmetry
respectively. Equation~\ref{AsyLorentz} is a ``truncated" form (i.e
the non-linear terms have been removed from the numerator) of the
commonly employed asymmetric expression of \cite{Nigam1998}. The
reason for using this truncated form is to have an expression that
remains physically reasonable far away from the central frequency of
the mode, which, as will be explained shortly, is an important
factor for the PGM. This truncated expression also has a physical
significance as it matches the profile one derives assuming that
asymmetry is based solely on correlations with a flat background
(see \citealt{Toutain2006}).

The model was fitted to the data using an appropriate
maximum-likelihood estimator \citep{Anderson1990}. All visible modes
within the frequency ranges 1300 $ \leq \nu \leq $ 4600 $\mu$Hz and
with angular degree 0 $ \leq \ell \leq $ 3 were included. For much
of the frequency range the weak $\ell$=4 and 5 modes fall within the
range of the even-$\ell$ fitting windows. Therefore, in the region
of the spectrum where the modes are strongest, the fitting model
also included parameters for the $\ell$ = 4 and 5 modes. It has
previously been shown that, despite their low visibility, if the
$\ell$ = 4 and 5 modes are not accounted for they can often impact
on the fitted parameters of the stronger modes (see
\citealt{Chaplin2006,Fletcher2007,Jimenez2008}). The even-$\ell$
windows were set at a size in frequency of 65.1 $\mu$Hz when the
$\ell$ = 4 and 5 modes were included and at 44.2 $\mu$Hz when they
were not (i.e., at low and high frequencies were the $\ell$ = 4 and
5 modes have extremely low amplitudes). The odd-$\ell$ windows had a
size of 53.6 $\mu$Hz in frequency. The mode and background
parameters described in Paper I were varied iteratively until they
converged upon their best-fitting values.

\begin{figure}
\centerline{\includegraphics[width=3.4in]{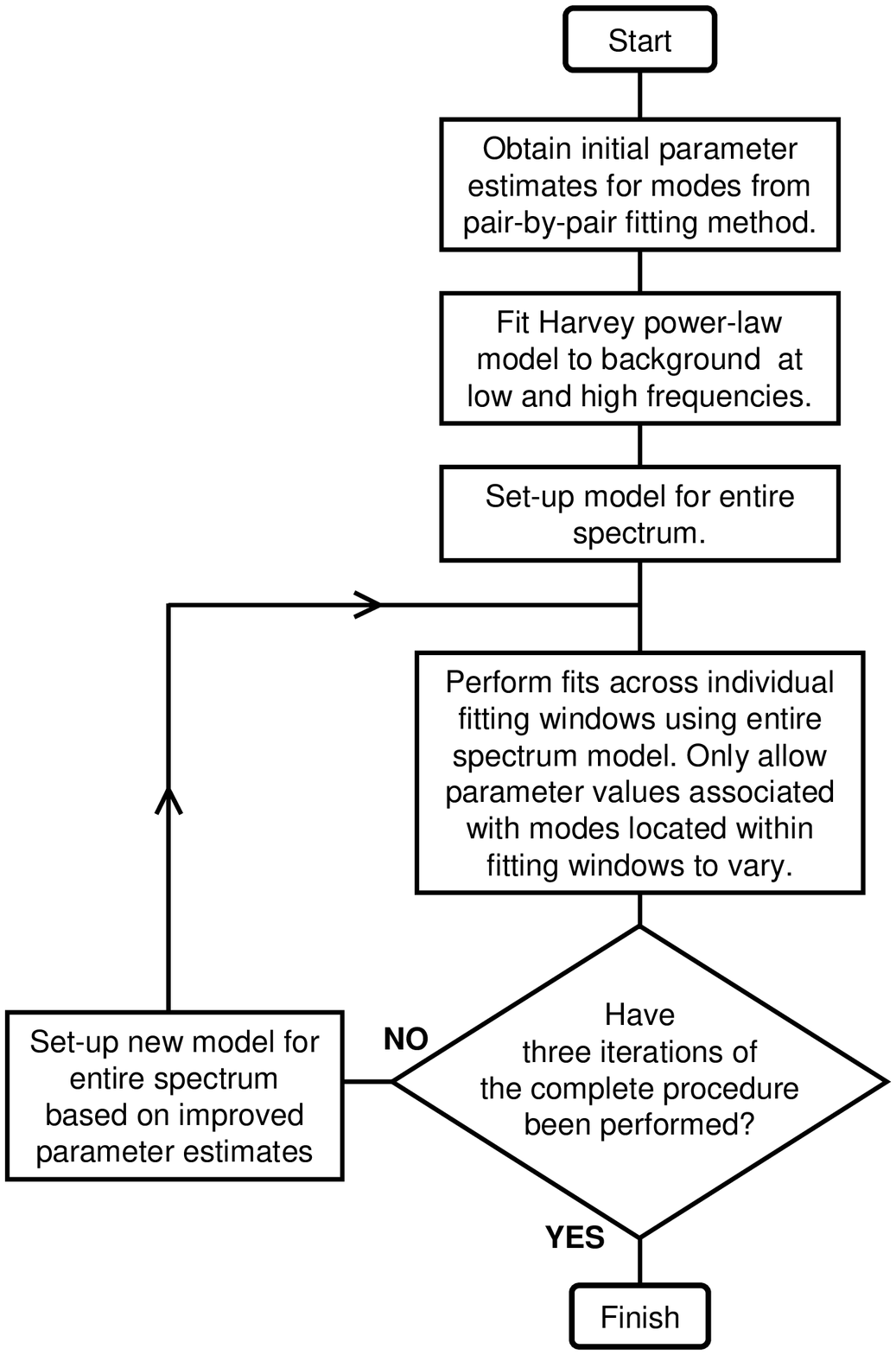}}\caption{Step-by-step
flow diagram representing the PGM fitting algorithm.}
\label{PG_FlowChart}
\end{figure}

We now go on to describe the additional steps needed for the PGM, a
step-by-step flow diagram of which is given in
Fig.~\ref{PG_FlowChart}. The optimisation is still carried out in
fitting windows with only the parameters of the modes that lie
within these regions being allowed to vary. Now, however, the
starting model for the fit is generated for the entire spectrum,
derived from the results of an initial run of the PPM method. In
this way we are accounting for any extra power arising from the
wings of modes whose central frequencies lie outside the fitting
window but are still optimising only a relatively small number of
parameters at any one time.

There are three refinements made to the PGM used here compared to
the method outlined in Paper I. The first is how we deal with
asymmetries. In Paper I the simulated data contained only symmetric
modes and so asymmetries were set to zero when forming the initial
estimate of the full spectrum model (i.e., the impact of modes lying
outside the fitting windows was based on symmetric profiles). In
this work the asymmetries returned by the PPM were retained and the
initial estimate of the full spectrum model was built from
asymmetric modes described by Equation~\ref{AsyLorentz}. The fact
that non-zero asymmetries are used in creating the full spectrum
model for the PGM is the reason why an expression such as
Equation~\ref{AsyLorentz}, which remains physically reasonable well
away from the central frequencies of the modes, is needed.

The same parameters were varied in the PGM stage as were fitted in
the PPM stage. However, in order to reduce the complexity of the
code the sizes of the fitting windows were increased to include both
an even-$\ell$ and an odd-$\ell$ mode pair (and consequently the
corresponding $\ell$ = 4 and 5 modes as well). A window size of
130.1 $\mu$Hz was chosen to cover this range of modes. Since more
modes were being fitted within each window, a larger number of
fitting parameters was needed, however the extra computing time
needed because of this is partially offset by needing fewer actual
fitting windows to cover the whole spectrum.

The increased window size also meant it became more important to
take account of the non-white background profile. In Paper I, a
varying background that roughly mimics the granulation profile seen
in real data was achieved by fitting a background parameter divided
by the frequency. This time a different approach was employed.
Before fitting the modes, the background was fitted at low and high
frequencies, well away from the regime of the resonant modes. This
fit was performed using the \cite{Harvey1985} power-law model with
an additional constant offset term to account for photon shot noise
as described in $\S$~\ref{SecData}. From this fit the background was
interpolated over the entire spectrum. Therefore, during the mode
fitting stage, the background will have a non-white form as
described by the fitted Harvey Model. However, in order to reduce
the number of parameters only the constant offset term was allowed
to vary during the mode fitting stage.

The final difference between the PGM given here and that introduced
in Paper I comes in performing an extra iterative process in order
to further refine the parameter estimates. By employing the PGM a
more accurate set of parameters is returned compared with those
estimated from the PPM. This new set of parameters was then used to
set-up a more accurate full spectrum model. The entire PGM process
can then be performed again to produce a further set of parameters.
Tests with simulated data have shown that performing more than three
iterations of this process does not bring any further significant
changes in the new parameter set compared with the previous one (the
difference between the fitted parameters returned from the third
iteration compared with those from the second was almost always less
that 0.1 sigma). Based on this we chose to perform only three
iterations of the procedure in all our fits to the simulated and
real data.

In terms of efficiency, the time taken to complete one complete
iteration for the entire spectrum using the PGM is about three times
longer than that of the PPM. Assuming three additional complete
iterations are needed to produce the best fitting parameters, the
overall computing time will be of the order of 10 to 15 times
longer. Nevertheless, it still takes significantly less time than
that which would be needed to perform a full global spectrum fit.

\section{Results}\label{SecResults}

The main goal of this work is to develop a better, but still
computationally efficient, peak-bagging technique for real
helioseismic data, and thus enable more accurate inferences about
the Sun. However, in order to validate the new method, we first give
the results of tests carried out on simulated data.

\subsection{Simulated Data}\label{SecSimFits}

The PPM and PGM outlined in this paper were used to fit two
simulated 796-day limit spectra with GOLF-like background
characteristics. It should be noted that in all the plots that
follow we only give the results up to around 4000 $\mu$Hz. While we
do fit at higher frequencies the parameter estimates for these modes
tend to have very large uncertainties due to their short lifetimes
(i.e., large linewidths). Hence, above 4000 $\mu$Hz, any improvement
in the accuracy of the fits due to using the PGM is unlikely to be
significant.

\begin{figure*}
\centerline{\includegraphics[width=2.8in]{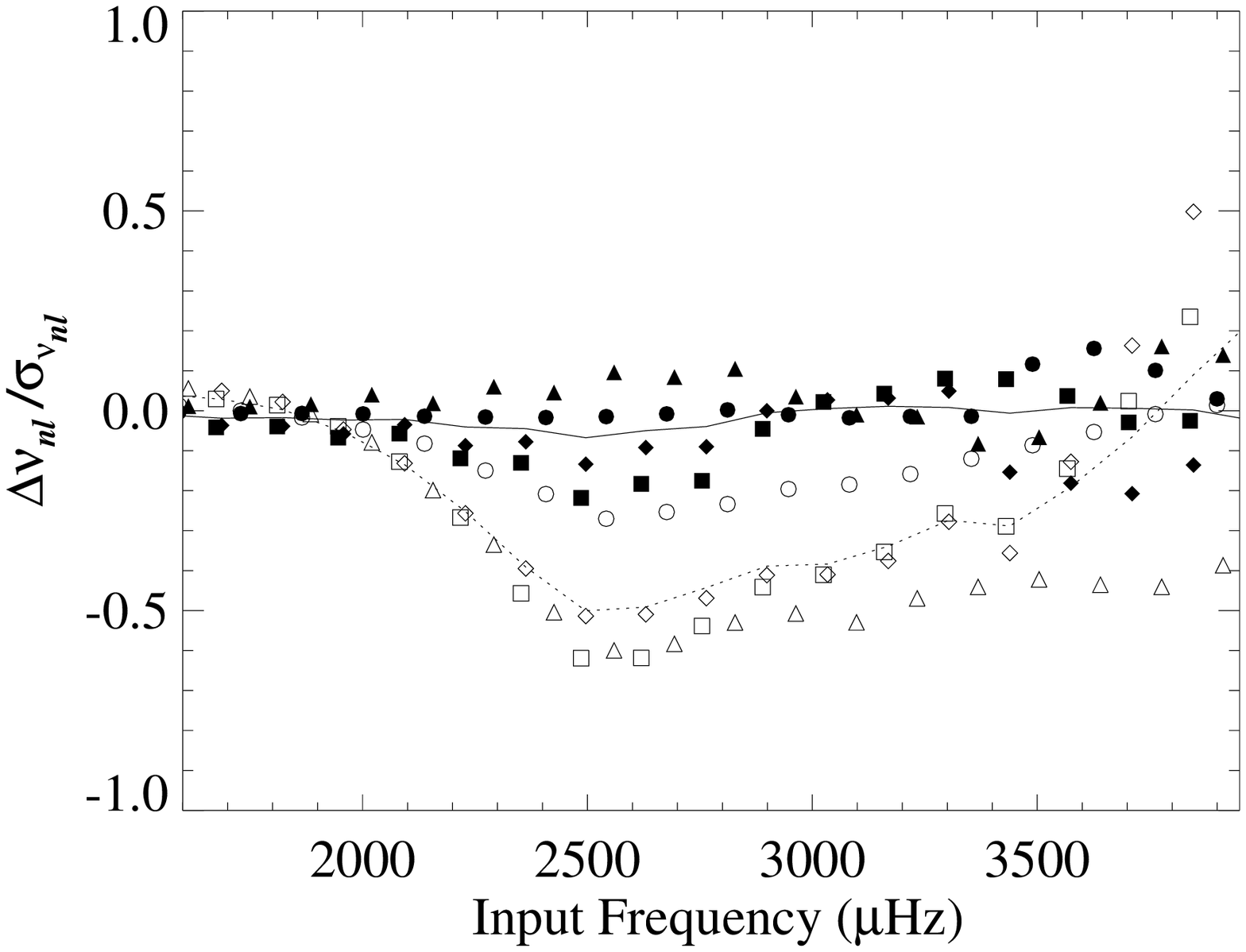}
\includegraphics[width=2.8in]{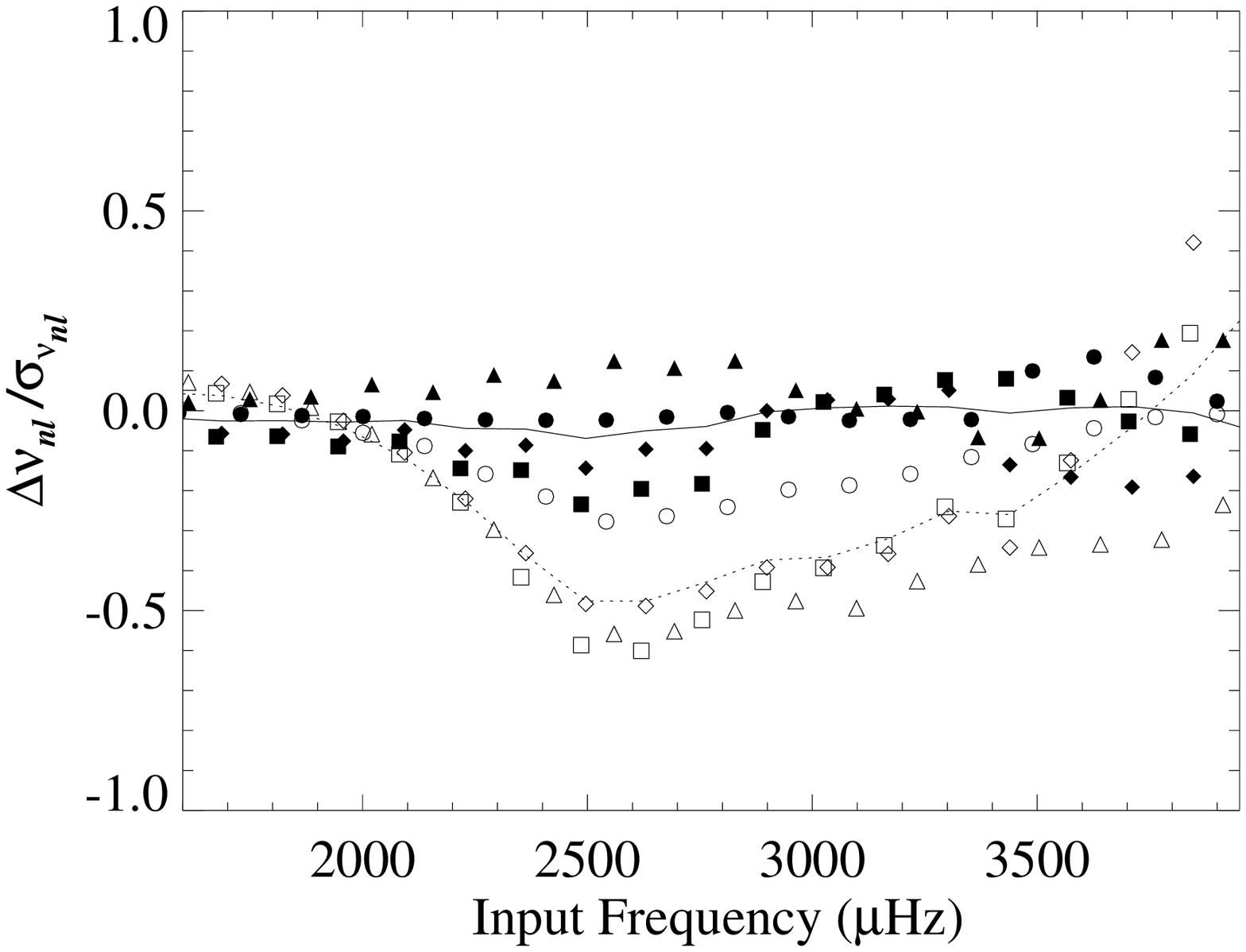}}
\caption{Differences between the fitted and input frequencies (in
the sense fitted minus input) divided by the formal uncertainties
when fitting simulated limit spectra. The results for the limit
spectrum created by correlating the modes with a flat background are
shown in the left panel while the results for the limit spectrum
created by correlating the modes with a granulation-like background
are shown in the right panel. Open symbols give the estimates
returned by the PPM while solid symbols give estimates returned by
the PGM. Diamonds signify $\ell$ = 0 modes, triangles $\ell$ = 1,
squares $\ell$ = 2 and circles $\ell$ = 3. The solid and dotted
lines give the differences averaged over four consecutive modes in
frequency (i.e., one mode each of $\ell$ = 0, 1, 2 and 3).}
\label{SimFreq}
\end{figure*}

\begin{figure*}
\centerline{\includegraphics[width=2.8in]{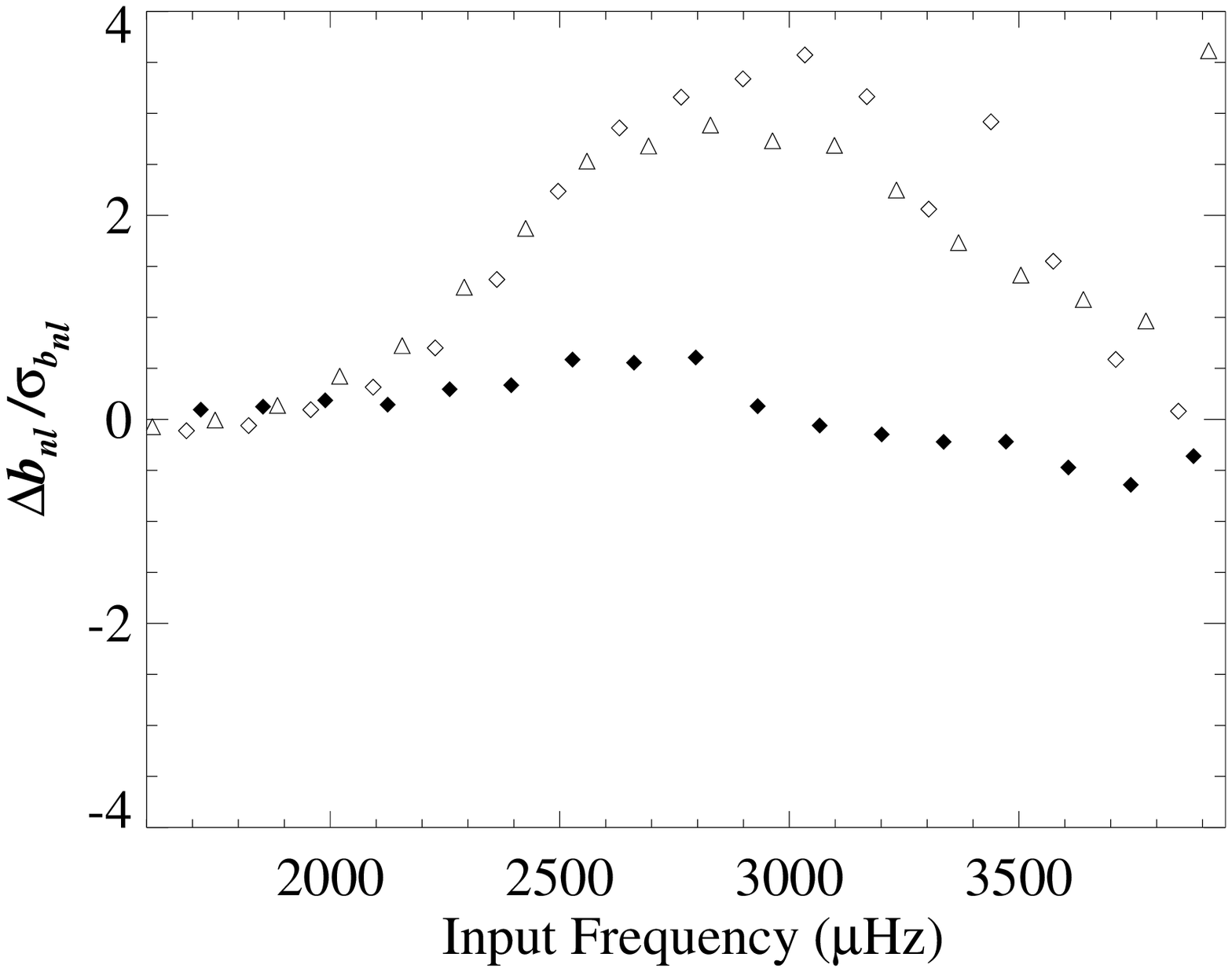}
\includegraphics[width=2.8in]{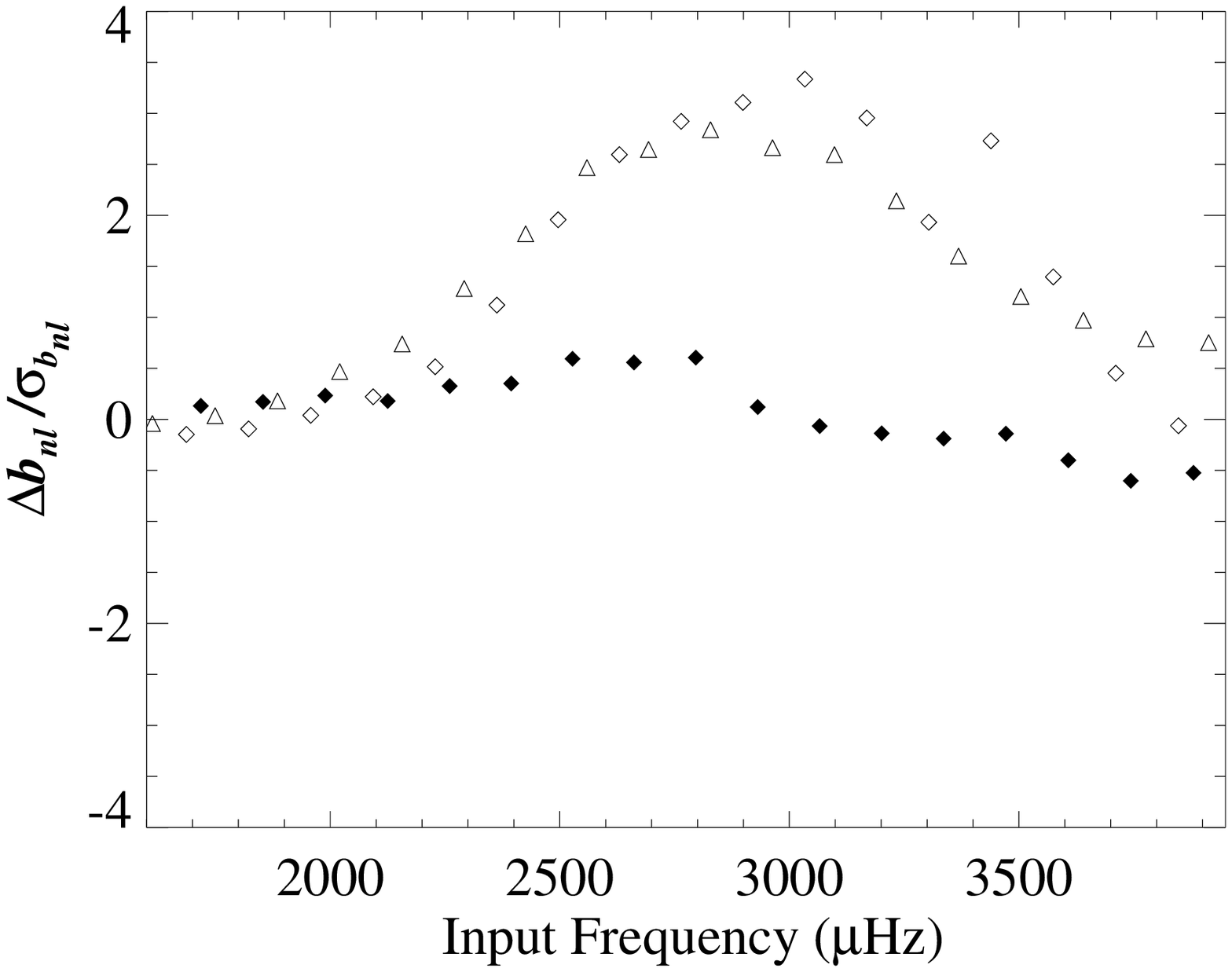}}
\caption{Differences between the fitted and input asymmetries (in
the sense fitted minus input) divided by the formal uncertainties
when fitting simulated limit spectra. The results for the limit
spectrum created by correlating the modes with a flat background are
shown in the left panel while the results for the limit spectrum
created by correlating the modes with a granulation-like background
are shown in the right panel. Open symbols give the results returned
by the PPM with diamonds giving the estimates from $\ell$ = even
windows and triangles $\ell$ = odd windows. Solid symbols give the
results from the PGM across each extended fitting window (see
text).} \label{SimAsym}
\end{figure*}

\begin{figure*}
\centerline{\includegraphics[width=2.8in]{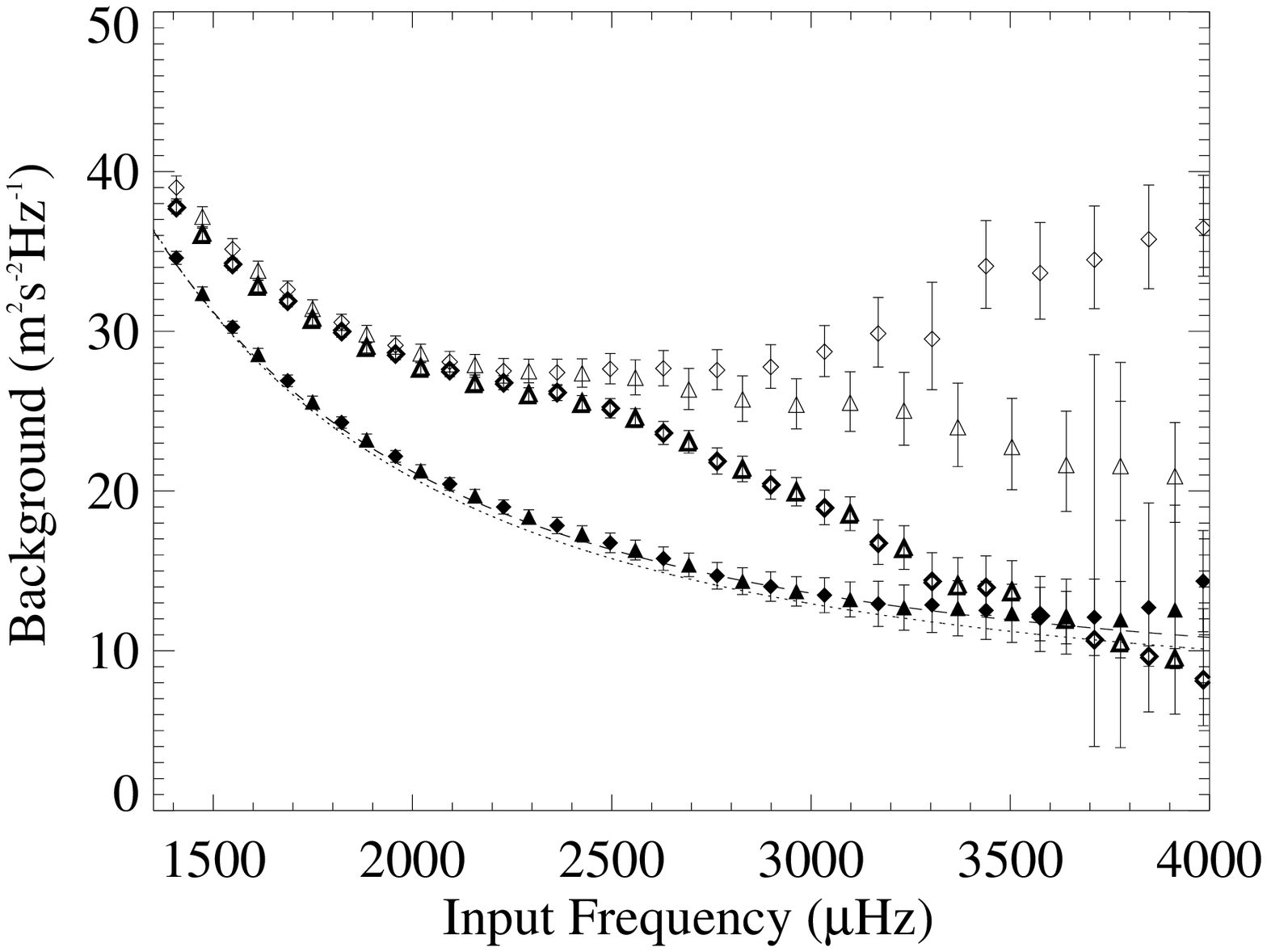}
\includegraphics[width=2.8in]{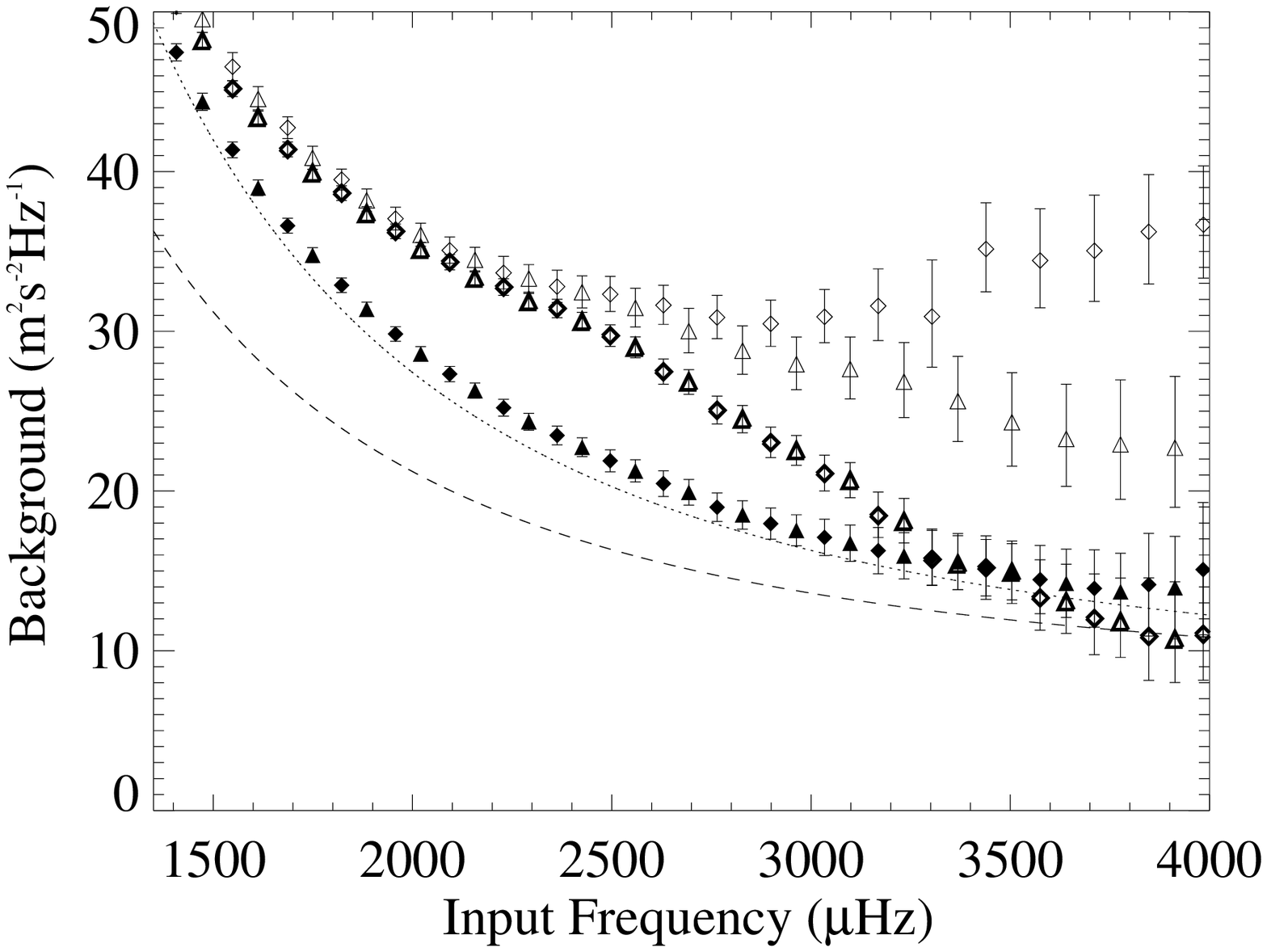}}
\caption{Estimated backgrounds when fitting simulated limit spectra.
The results for the limit spectrum created by correlating the modes
with a flat background are shown in the left panel while the results
for the limit spectrum created by correlating the modes with a
granulation-like background are shown in the right panel. Open
symbols give the results returned by the PPM, bold open symbols give
the results from the PGM outlined in Paper I, and solid symbols give
the results from the PGM outlined in this paper. Estimates of the
background at $\ell$ = 0 frequencies are given by diamonds and at
$\ell$ = 1 frequencies by triangles. The dashed lines give the real
input backgrounds and the dotted lines give the interpolated
backgrounds.} \label{SimBG}
\end{figure*}

Fig.~\ref{SimFreq} shows the fitted estimates returned for the
central frequency parameter, $\nu_{nl}$. In order to give a direct
measurement of the significance of any bias in the results we have
plotted the difference between the fitted values and the known input
values (in the sense fitted - input) and divided by the estimated
uncertainties, $\sigma_{\nu_{nl}}$. The uncertainties were
calculated by taking the square root of the diagonal elements of the
inverted Hessian fitting matrix. In this figure, and those that
follow, different symbols have been used for different $\ell$, and
open and solid symbols have been used for the PPM and PGM results
respectively. Also, for the figures showing simulated data
(Figs.~\ref{SimFreq} to \ref{SimBG}) we have given plots for both
the flat correlated background data (left panel) and the
granulation-like correlated background data (right panel).

In both panels the plots show that there is a clear systematic bias
seen in the fitted frequencies returned by the PPM. This result can
be explained by the fact that asymmetric modes lying outside a
particular fitting window will introduce a different level of excess
power at the low frequency end of the fitting window compared with
the high frequency end. Since this is not accounted for by the PPM,
the fitted asymmetries will be biased and consequently affect the
fitted frequencies.

The fact that fitting symmetric Lorentzian models to asymmetric
modes leads to biased frequencies is well known in the literature
(e.g., see \citealt{Chaplin1999,Antia1999}) and so it is not
surprising that inaccurate asymmetries will also lead to biased
frequencies. A similar bias was seen in the frequency estimates
returned in a solarFLAG ``hare-and hounds" exercise when asymmetric
modes were fitted, as reported in \cite{Jimenez2008}. However, in
Paper I, where symmetric modes were fitted, it was shown that the
PPM returned robust estimates of the frequencies. This adds further
evidence that it is the inaccuracy in the asymmetries returned by
the PPM that results in the biased frequencies.

Fig.~\ref{SimFreq} also shows that the bias in the frequencies
returned by the PGM is significantly reduced. This is a result of
the bias in the asymmetries being reduced by accounting for the
offset excess power. The small amount of bias that still remains in
the fitted frequencies returned by the PGM is strongly dependent on
the angular degree of the modes. This is due to the fact that only a
single asymmetry parameter was fitted to all the modes within a
fitting window, whereas the input asymmetries actually differ
slightly from mode to mode. Finally, a comparison of the two panels
in Fig.~\ref{SimFreq} shows that the frequency estimates are very
similar whichever background type is used.

The expected bias in the asymmetry parameter returned from the PPM
is shown in Fig.~\ref{SimAsym}. Again the results are plotted as the
difference between the fitted and input values and divided by the
uncertainties. As the asymmetries are assumed to be constant across
all modes within each fitting window, we only plot one value per
window. In the case of the PPM we have separate results for the
$\ell$ = odd and even mode pairs, while for the PGM, where the
fitting windows were increased to encompass both pairs, we only plot
a single value.

A clear improvement in the fitted asymmetries returned from the PGM
can be seen. For the PPM the fitted asymmetries are significantly
greater than the input values by as much as 3 sigma. Since the
asymmetries are negative, this actually means the PPM returns less
negative values. In effect this means that the magnitudes of the
fitted asymmetries are smaller than the magnitude of the input
values. For the PGM any bias is consistently less than 1 sigma.
Again, the results are very similar for both the flat background
correlated data and the granulation-like background correlated data.

Fig.~\ref{SimBG} shows the fitted backgrounds. Results are again
shown for both the flat background correlated spectrum and the
granulation-like background correlated spectrum. In these plots, in
addition to showing the fits from both fitting strategies described
in this paper, we also include the results of fitting the less
sophisticated PGM outlined in Paper I (hereafter abbreviated OPGM).

The plots show how the background fits from the PPM overestimate the
true background to a large extent. The OPGM improves upon this, but,
due to not allowing for the asymmetric nature of the modes lying
outside the fitting windows, a significant bias still remains. Only
when using the PGM outlined in this paper are estimates much closer
to the input background.

When fitting the flat background correlated spectrum, the PGM
reproduces the input background very well (given by the smooth
dashed lines in the plots). In the case where the input parameters
are unknown, one can still make an estimate of the background in the
vicinity of the modes by fitting a sensible model to low and high
frequencies and interpolating (this interpolated background is given
by a smooth dotted line in the plots). When a one-component Harvey
model with a constant offset term to account for photon shot-noise
was used to fit the flat background correlated data in this way, a
very close estimate of the input was obtained.

In contrast, when the excitation functions were correlated with a
granulation-like background, background fits from the PGM, while
still reproducing the interpolated background well, significantly
overestimated the true background. This is because the excitation
function of the modes is correlated with a non-white frequency
response, and as such the mode profile will increase towards low
frequencies in the same manner as the background profile. This
introduces an extra ``pedestal" of power into the spectrum which is
not accounted for by the fitting model. This pedestal also adds to
the background power at low frequencies and as such will impact the
interpolated background value as well.

\subsection{Real Data}\label{SecRealFits}

\begin{figure*}
\centerline{\includegraphics[width=2.8in]{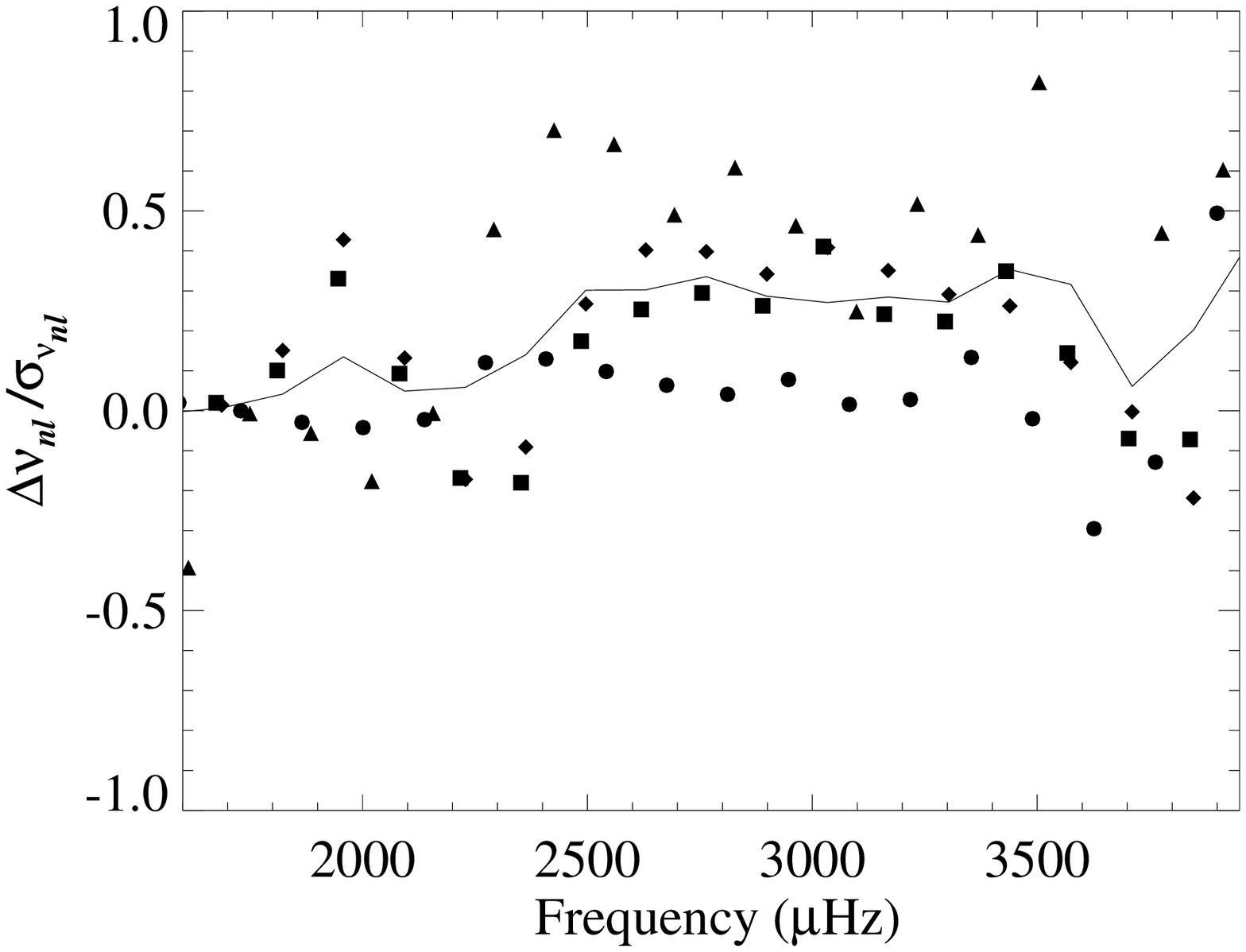}
\includegraphics[width=2.8in]{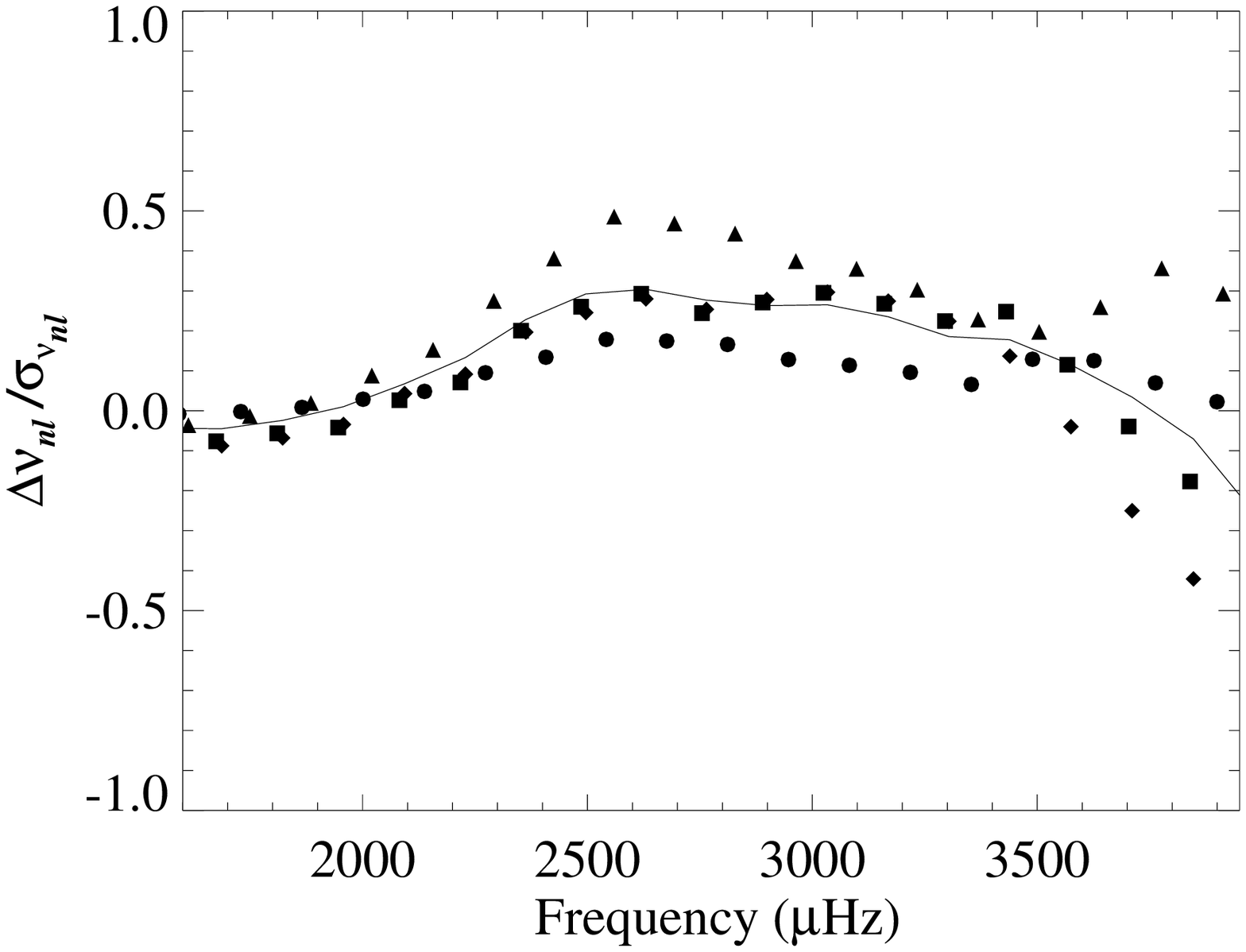}}
\caption{Left panel: Differences between the fitted frequencies
returned by the PPM and PGM routines (in the sense PGM minus PPM)
divided by the combined formal uncertainties when fitting a spectrum
made from a real 796-day time series of GOLF observations. Right
panel: The same as in the left panel but for the simulated
granulation-like noise excited data. Diamonds signify $\ell$ = 0
modes, triangles $\ell$ = 1, squares $\ell$ = 2 and circles $\ell$ =
3. The solid line gives the differences averaged over four
consecutive modes in frequency (i.e., one mode each of $\ell$ = 0,
1, 2 and 3).} \label{RealFreq}
\end{figure*}

The PPM and PGM were also used to fit a real set of GOLF data.
Although in this case we cannot test the fitted values against known
input parameters, we can plot the differences between the results
from the two fitting methods. This is shown for the fitted
frequencies in the left panel of Fig.~\ref{RealFreq} (in the sense
PGM minus PPM). We have again divided the differences by the
(combined) formal uncertainties, in order to show more clearly the
significance of the differences.

The plot shows that the fitted frequencies are higher when using the
PGM. Also the differences appear to be largest for $\ell$ = 1 modes
and smallest for $\ell$ = 3 modes. This matches what was seen when
fitting the simulated limit spectrum as shown in the right panel of
Fig.~\ref{RealFreq} where we have replotted the simulated results as
PGM minus PPM. As with the simulated data it is likely that the
different frequency estimates are a consequence of the two
techniques returning different asymmetries.

\begin{figure*}
\centerline{\includegraphics[width=2.8in]{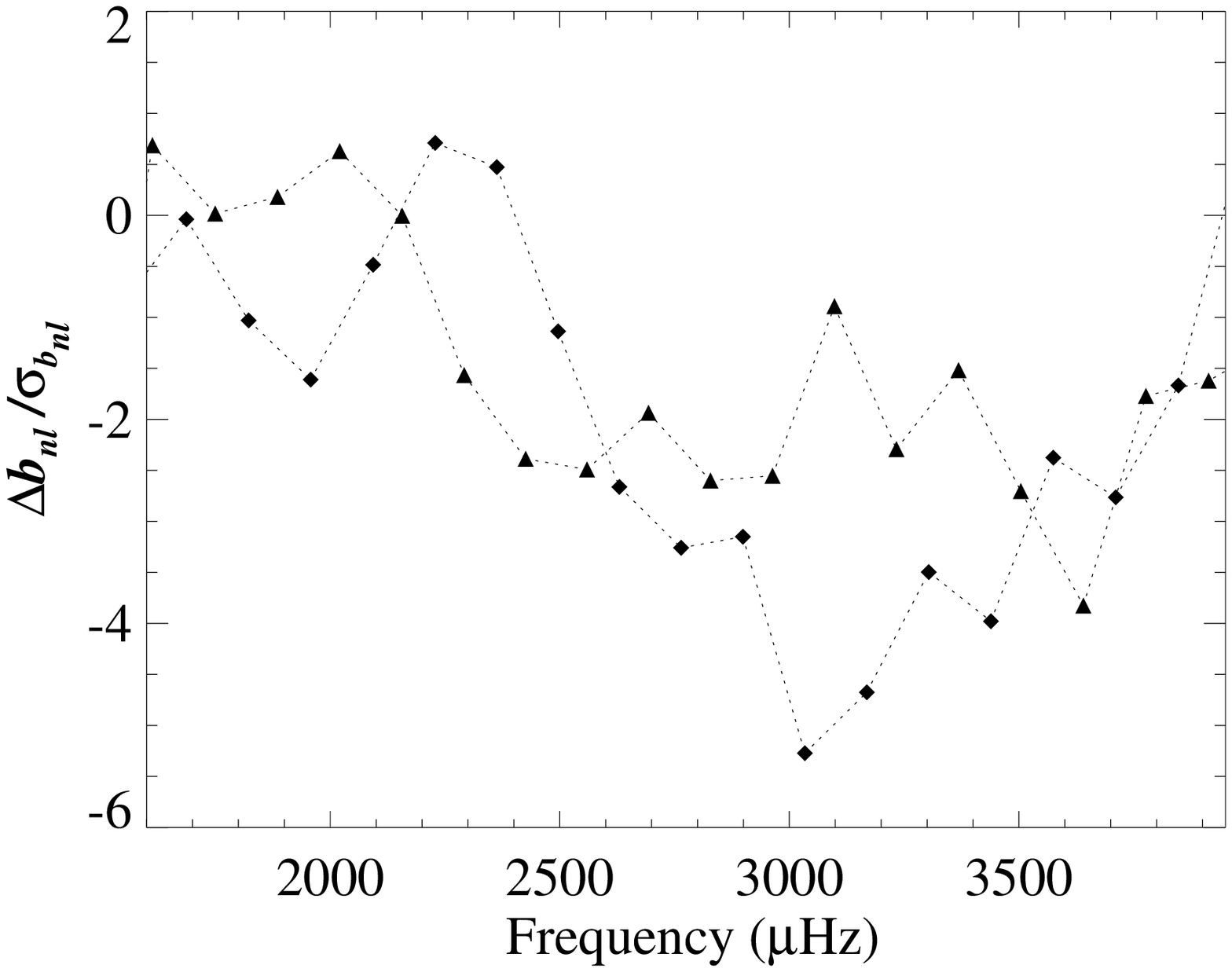}
\includegraphics[width=2.8in]{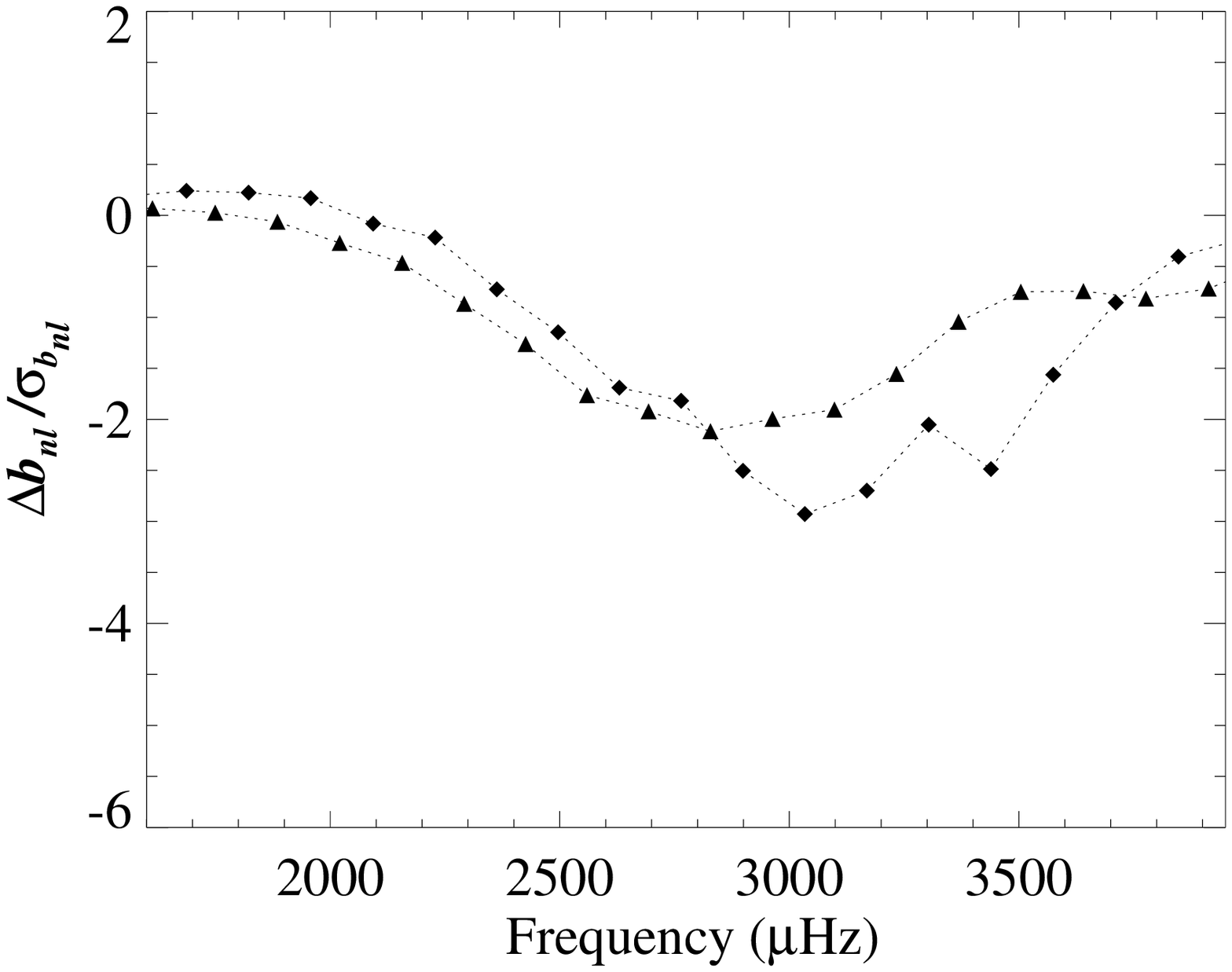}}
\caption{Left panel: Differences between the fitted asymmetries
returned by the PPM and PGM routines (in the sense PGM minus PPM)
divided by the combined formal uncertainties when fitting a spectrum
made from a real 796-day time series of GOLF observations. Right
panel: The same as in the left panel but for the simulated
granulation-like noise excited data. Diamonds signify the
differences between asymmetry fits in the $\ell$ = even windows
compared with the PGM fits and triangles signify the differences
between asymmetry fits in the $\ell$ = odd windows and the PGM fits.
A dotted line joins the points to aid the eye.} \label{RealAsym}
\end{figure*}

Fig.~\ref{RealAsym} shows the difference in fits for the asymmetry
parameter. Again we see a bias between the results returned by the
PGM compared with those from the PPM. The estimates from the PGM are
systematically and significantly more negative. This again matches
what was seen when fitting the simulated limit spectrum as shown in
the right panel of Fig.~\ref{RealAsym} where, as per the frequency
plots, we have replotted the simulated data in the form PGM minus
PPM.

\begin{figure*}
\centerline{\includegraphics[width=2.9in]{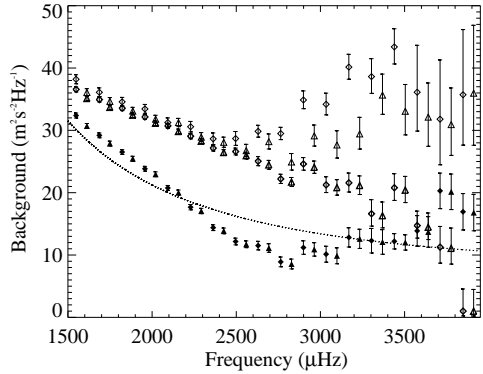}}\caption{Estimated
backgrounds when fitting a spectrum made from a real 796-day time
series of GOLF observations. Open symbols give the results returned
by the PPM, bold open symbols give the results from the PGM outlined
in Paper I, and solid symbols give the results from the PGM outlined
in this paper. Estimates of the background at $\ell$ = 0 frequencies
are given by diamonds and at $\ell$ = 1 frequencies by triangles.
The dotted line gives the interpolated background.} \label{RealBG}
\end{figure*}

Finally, Fig~\ref{RealBG} shows a plot of the fitted backgrounds. We
again include fits returned from the less sophisticated PGM
described in Paper I. As mentioned in section~\ref{SecSimFits} we
can compare the fitted backgrounds with the background interpolated
from fitting at low and high frequencies.

The fits from the PGM outlined in this paper clearly match the
interpolated background better than the other two methods. However,
the match is not perfect. Between about 2200 - 3100 $\mu$Hz the
fitted backgrounds significantly underestimate the interpolated
background wheras above 3500 $\mu$Hz there is an overestimation. The
most likely source of error is in modeling the mode asymmetry, since
incorrectly fitted asymmetries can have an effect on the background
as seen in Fig~\ref{RealBG}.

\section{Summary}\label{SecSummary}

We have further developed a new ``pseudo-global" fitting method
(PGM) designed to gain the advantages of fitting simultaneously the
entire Sun-as-a-star p-mode spectrum while retaining computational
efficiency. By fitting the limit spectra of simulated data it was
shown that this new method enabled more accurate estimates of mode
frequencies and asymmetries to be returned when compared with the
estimates from a solely ``local" pair-by-pair fitting method (PPM).

When fitting a real set of GOLF data it was shown that there was a
systematic difference between the fitted asymmetries, and
consequently the fitted frequencies, returned by the PPM and PGM. It
will be the subject of further work to test whether or not such
systematic biases in the frequency estimates of the low-degree modes
have any significant effect on inversions to infer the solar
structure. Even so, from the simulated data, we would expect that
the bias in the fitted frequencies returned from the PGM to be very
small. As such we think it would be of benefit to employ the PGM in
future analysis of Sun-as-a-star data.

Fits to the simulated data showed that, in the case of the
excitation profiles of the modes being correlated with a flat
background, the new PGM returned estimates of the background very
close to the input values. It is also possible to estimate the
background by fitting an appropriate model at low and high
frequencies and then interpolating for the frequency range where the
resonant modes occur. For the flat background correlated data this
``interpolated" value was also shown to match the input values
reasonably well. However, in the case of the excitation profiles
being correlated with a granulation-like background, both the fitted
values and the interpolated values overestimated the true
background.

In the real GOLF data the PGM gave estimates of the background that
matched much more closely the interpolated background than those
from the PPM. However, although the simulated data showed there is a
difference in the returned backgrounds for the two different source
types used, this only shows up in comparison with the input
background. As this is unknown for the real data we cannot easily
distinguish whether the modes are excited via correlations with a
flat background, correlations with a granulation-like background or
some mixture of the two. This also means we cannot be sure whether
the asymmetric profile as used in this paper (i.e., asymmetry
produced solely via correlation between the modes and the
background) is physically a good match for the real modes. However,
the small but significant differences between the fitted background
compared with the interpolated background in the regions 2200 - 3100
$\mu$Hz and above 3500 $\mu$Hz suggest we do not yet have a perfect
understanding of how the modes are generated. This could be due to a
lack of understanding of how the granulation-like noise behaves in
the vicinity of the modes or an incomplete understanding of how
asymmetric modes are generated.

At the time of writing, the PGM had been successfully adapted in
order to fit gapped data. We therefore intend to employ the
technique to fit BiSON data where the time series length available
is of the order of 30 years.

\section*{Acknowledgments}

STF acknowledges the support of the School of Physics and Astronomy
at the University of Birmingham, the Faculty of Arts, Computing,
Engineering and Science (ACES) at the University of Sheffield and
the support of the Science and Technology Facilities Council (STFC).
We also thank all those associated with BiSON which is funded by the
STFC. We are grateful to all those involved in making the GOLF data
publicly available especially R. Garc\'{\i}a for calibrating and
preparing the particular data sets used in this paper. SOHO is a
mission of international cooperation between ESA and NASA.

\bibliography{fletcher}

\end{document}